\documentclass[english,fleqn]{llncs}
\usepackage{amsmath} 
\usepackage{mystyle,mymath,proof,latexsym,amssymb}
\usepackage{color}
\usepackage{graphicx}

\def\T#1{\hbox{\color{green}{$\clubsuit #1$}}}

\begin{document}
\sloppy \hbadness=10000 \vbadness=10000

\def\prooflineskip{\def\arraystretch{2.5}}


\def\Arr{{\rm Arr}}


\def\In{{\rm In}}
\def\Tri#1#2#3{\{{#1}\}\hskip 0.5ex{\tt {#2}}\hskip 0.5ex\{{#3}\}}
\def\Loc{{\rm Loc}}
\def\Verify{{\rm Verify}}
\def\Norm{{\rm Norm}}
\def\Malloc{{\tt malloc}}
\def\Free{{\rm free}}
\def\Fail{{\rm fail}}
\def\Simp{{\rm Simp}}
\def\Abst{{\rm Abst}}
\def\Any{{\rm Any}}
\def\Eval#1{\llbracket{#1}\rrbracket}


\def\H#1#2{{\hat h|_{{#1}(\Roots({#2}))}}}
\def\Array{{\rm Arr}}
\def\Perm{{\rm Perm}}
\def\Sorted{{\rm Sorted}}
\def\mkPi{{\rm mkPi}}

\def\Length{{\rm length}}
\def\Nth{{\rm nth}}
\def\Replace{{\rm replace}}


\def\SLID{{\rm SLID}}
\def\CSLID{{\rm CSLID}}
\def\Eclass{{\rm Eclass}}
\def\Eq{{\rm Eq}}
\def\Deq{{\rm Deq}}
\def\Satom{{\rm Satom}}
\def\Cells{{\rm Cells}}
\def\Roots{{\rm Roots}}
\def\Elim{{\rm Elim}}
\def\Case{{\rm Case}}
\def\Unfold{{\rm Unfold}}
\def\Pred{{\rm Pred}}
\def\Start{{\rm Start}}
\def\Unsat{{\rm Unsat}}
\def\Split{{\rm Split}}
\def\Underscore{\underline{\phantom{x}}}
\def\Rootcell{{\rm rootcell}}
\def\Rootshape{{\rm rootshape}}
\def\Jointleaf{{\rm jointleaf}}
\def\Jointleafimplicit{{\rm jointleafimplicit}}
\def\Jointnode{{\rm jointnode}}
\def\Directjoint{{\rm directjoint}}


\def\Range{{\rm Range}}
\def\Xstart{X_{{\rm start}}}
\def\kstart{k_{{\rm start}}}
\def\Leaves{{\rm Leaves}}
\def\PureDist{{\rm PureDist}}
\def\Pure{{\rm Pure}}
\def\Identity{{\rm Identity}}
\def\Subst{{\rm Subst}}


\def\Connected{{\rm Connected}}
\def\Eststablished{{\rm Eststablished}}
\def\Valued{{\rm Valued}}


\def\Nil{{{\rm nil}}}
\def\eqDef{=_{{\rm def}}}
\def\Inf#1{{\infty_{#1}}}
\def\Stores{{\rm Stores}}
\def\SVars{{{\rm SVars}}}
\def\Val{{{\rm Val}}}
\def\MSO{{{\rm MSO}}}
\def\Sep{{{\rm sep}}}
\def\THeaps{{\rm THeaps}}
\def\Root{{\rm Root}}
\def\TG{{\rm TGraph}}

\def\Tilde{\widetilde}
\def\Bar{\overline}
\def\Lequiv{\Longleftrightarrow}
\def\Lto{\Longrightarrow}
\def\Lfrom{\Longleftarrow}
\def\Noshare{{\rm Noshare}}
\def\Roots{{\rm Roots}}
\def\Forest{{\rm Forest}}
\def\Var{{\rm Var}}
\def\Range{{\rm Range}}
\def\Cell{{\rm Cell}}
\def\Tree{{\rm Tree}}
\def\Switch{{\rm Switch}}
\def\All{{\rm All}}
\def\To{\leadsto}
\def\tree{{\rm tree}}
\def\Paths{{\rm Paths}}
\def\Finite{{\rm Finite}}
\def\Const{{\rm Const}}
\def\Leaf{{\rm Leaf}}
\def\LeastElem{{\rm LeastElem}}
\def\LeastIndex{{\rm LeastIndex}}
\def\WSnS{{{\rm WSnS}}}
\def\Expand{{{\rm Expand}}}


\long\def\J#1{} 
\def\T#1{{\rm \color{green}{$\clubsuit #1$}}}
\def\W#1{{\rm \color{Orange}{$\spadesuit #1$}}}

\def\Node{{{\rm Node}}}
\def\LL{{{\rm LL}}}
\def\DSN{{{\rm DSN}}}
\def\DCL{{{\rm DCL}}}
\def\LS{{{\rm LS}}}
\def\Ls{{{\rm ls}}}

\def\FPV{{{\rm FPV}}}
\def\Lfp{{{\rm lfp}}}
\def\IsHeap{{{\rm IsHeap}}}

\def\Equiv{\quad \equiv\quad }
\def\Null{{{\rm null}}}
\def\Emp{{{\rm emp}}}
\def\If{{{\rm if\ }}}
\def\Then{{{\rm \ then\ }}}
\def\Else{{{\rm \ else\ }}}
\def\While{{{\rm while\ }}}
\def\Do{{{\rm \ do\ }}}
\def\Cons{{{\rm cons}}}
\def\Dispose{{{\rm dispose}}}
\def\Vars{{{\rm Vars}}}
\def\Locs{{{\rm Locs}}}
\def\States{{{\rm States}}}
\def\Heaps{{{\rm Heaps}}}
\def\FV{{{\rm FV}}}
\def\True{{{\rm true}}}
\def\False{{{\rm false}}}
\def\Dom{{{\rm Dom}}}
\def\Abort{{{\rm abort}}}
\def\New{{{\rm New}}}
\def\W{{{\rm W}}}
\def\Pair{{{\rm Pair}}}
\def\Lh{{{\rm Lh}}}
\def\lh{{{\rm lh}}}
\def\Elem{{{\rm Elem}}}
\def\EEval{{{\rm EEval}}}
\def\PEval{{{\rm PEval}}}
\def\HEval{{{\rm HEval}}}
\def\Domain{{{\rm Domain}}}
\def\Exec{{{\rm Exec}}}
\def\Store{{{\rm Store}}}
\def\Heap{{{\rm Heap}}}
\def\Storecode{{{\rm Storecode}}}
\def\Heapcode{{{\rm Heapcode}}}
\def\Lesslh{{{\rm Lesslh}}}
\def\Addseq{{{\rm Addseq}}}
\def\Separate{{{\rm Separate}}}
\def\Result{{{\rm Result}}}
\def\Lookup{{{\rm Lookup}}}
\def\ChangeStore{{{\rm ChangeStore}}}
\def\ChangeHeap{{{\rm ChangeHeap}}}
\def\Wand{\mathbin{\hbox{\hbox{---}$*$}}}
\def\Vec{\overrightarrow}

\def\Tilde{\widetilde}
\def\Break{\hfil\break\hbox{}}

\def\SLAR{\textbf{SLAR}}
\def\SLG{\textbf{G}}
\def\Checker{\textbf{SLar}}
\def\ZZZ{\textbf{Z3}}
\def\Disj{\hbox{Disj}}
\def\Vars{\hbox{Vars}}
\def\Loc{\hbox{Loc}}
\def\Val{\hbox{Val}}
\def\Proof{\noindent{\em Proof.}\quad}
\def\Base{\hbox{{\sf Base}}}
\def\SingleFrame#1{\hbox{{\sf SingleFrame}-$#1$}}
\def\SingleNFrame#1{\hbox{{\sf SingleNFrame}-$#1$}}
\def\Multi{\hbox{{\sf Multi}}}
\def\U{\hbox{\textbf{U}}}
\def\F{\hbox{\textbf{F}}}
\def\COND{(\star)}

\renewcommand{\theequation}{\alph{equation}}

\title{
Decision Procedure for \\
Entailment of Symbolic Heaps with Arrays
}
\author{
Daisuke Kimura
\inst{1}
\and
Makoto Tatsuta
\inst{2}
}
\institute{%
Toho University
\\
\email{kmr@is.sci.toho-u.ac.jp}
\and
National Institute of Informatics
\\
\email{tatsuta@nii.ac.jp}
}

\maketitle

\begin{abstract}
This paper gives a decision procedure for the validity of entailment
of symbolic heaps in separation logic with Presburger arithmetic and
arrays.  The correctness of the decision procedure is proved under the
condition that sizes of arrays in the succedent are not existentially
bound.  This condition is independent of the condition proposed by the
CADE-2017 paper by Brotherston et al, namely, one of them does not imply the
other.  For improving efficiency of the decision procedure, some
techniques are also presented.  The main idea of the decision
procedure is a novel translation of an entailment of symbolic
heaps into a formula in Presburger arithmetic, and
to combine it with an external SMT solver.  
This paper also gives experimental results by an implementation,
which shows that the decision procedure works efficiently enough to use.
\end{abstract}

\section{Introduction}

Separation logic can be used to verify/analyze heap-manipulating imperative programs with pointers,
and mainly it is successful for verify/analyze memory safety~\cite{OHearn05}. 
The aim of our paper is also memory safety. 
The advantage of separation logic is modularity by the frame rule, by
which we can independently verify/analyze each function that may
manipulate heaps~\cite{OHearn11}. 
The study in this paper goes along this line.

The final goal of our research is to develop a fully-automated program verifier of pointer programs based on separation logic. 
For this, this paper introduces a formal system of symbolic heap fragment of separation logic with arrays, shows decidability of its entailment problem, then gives an implementation (called $\Checker$) of our decision procedure, and finally discusses its improvement for efficiency. 

{\em Symbolic heaps} are formulas of separation logic in a simple form  
$\exists\Vec x(\Pi\land\Sigma)$. 
The pure part $\Pi$ describes properties of between terms (denoted by $t$, $u$), which represent memory addresses and values. 
The spatial part $\Sigma$ is a separating conjunction of the empty predicate $\Emp$, the points-to predicate $t\mapsto u$, and the array predicate $\Array(t,u)$. It represents some shape of heaps: 
$\Emp$ means the empty heap, $t\mapsto u$ means the single heap that uses only one address $t$ and the value at $t$ is $u$, $\Array(t,u)$ means the heap that contains only an array starting from $t$ ending at $u$, a separating conjunction $\Sigma_1*\Sigma_2$ means a heap which can be split into two disjoint sub-heaps that are represented by $\Sigma_1$ and $\Sigma_2$. 

In order to achieve our final goal, it is necessary to develop a solver for the entailment problem.
An {\em entailment} has the form $\phi\vdash\phi_1,\ldots,\phi_k$, where $\phi$ and $\phi_i$ are symbolic heaps.
It is said to be valid when $\phi\to \bigvee_i\phi_i$ is valid with respect to the usual heap model.
The {\em entailment problem} is the validity checking problem of given entailments.

In the literature, many researches for verification of pointer programs based on symbolic-heap systems have been done. 
In particular symbolic-heap systems with inductive predicates have been studied intensively~\cite{OHearn04,OHearn05,Kanovich14,Brotherston14,Cook11,Enea13,Enea14,Iosif13,Iosif14,Tatsuta15}. 
Berdine et al.~\cite{OHearn04,OHearn05} introduced the symbolic-heap system with hard-coded list and tree predicates, and showed decidability of its entailment problem. Iosif et al.~\cite{Iosif13,Iosif14} considered the system with general inductive predicates, and showed its decidability under the bounded tree-width condition. Tatsuta et al.~\cite{Tatsuta15} introduced the system with general monadic inductive predicates.

Array is one of the primitive data structures of pointer programs.
It is an important issue of verifying pointer programs to ensure that there is no buffer overflow.
So we need to consider the array structure as primitive.
However, as far as we know, there are two researches about symbolic-heap systems that have arrays
as primitive~\cite{OHearn06,Brotherston17}.
Calcagno et al.~\cite{OHearn06} studied shape analysis based on symbolic-heap system in the presence of pointer arithmetic.
Brotherston et al.~\cite{Brotherston17} investigated several problems about a symbolic-heap system with arrays. 

When we extend separation logic with arrays, it may be different from previous array
logics in the points that (1) it is specialized for memory safety, and
(2) it can scale up by modularity. 
Bradley et al.~\cite{Bradley06}, Bouajjani et al.~\cite{Bouajjani09}, and Lahiri et al.~\cite{Lahiri08} 
discussed logics for arrays but their systems are essentially
different from separation logic. We cannot apply their techniques to our
case.
Piskac et al.~\cite{Piskac13} proposed a separation logic system with list
segments, and it can be combined with various SMT solvers, including
array logics. However, if we combine it with array logics, the arrays
are external and the resulting system does not describe the arrays by
spatial formulas with separating conjunction. So their techniques
cannot solve our case.

In \cite{Brotherston17}, they proposed a decision procedure for the entailment problem by
giving an equivalent condition to existence of a counter-model for a given entailment, then 
checking a Presburger formula that expresses the condition. 
In order to do this, they imposed the restriction that the second argument of
a points-to predicate in the succedent of an entailment is not existentially bound. 

Our motivating example is $\Array(x,x) \vdash x \mapsto 0, \exists y(y > 0 \land x \mapsto y)$,
which means that
if the heap consists of the array from $x$ to $x$,
then the heap consists of a single cell at address $x$ with content $0$,
or
the heap consists of a single cell at address $x$ with content $y$ for
some $y>0$.
Since it is trivially true and simple, 
we expect an entailment checker could decide it.
So far there was no decision procedure of a class of entailments which contains this example,
since it does not satisfy the restriction of \cite{Brotherston17}.

The current paper shows decidability of the entailment problem under a condition:
the sizes of arrays in the succedent of the given entailment do not contain any existential variables.
It means that the shape of heaps represented by the succedent is completely determined by the antecedent. 
We need this condition for proving correctness of our decision procedure.
Our result decides an independent class of entailments (including our motivating example) to the class decided by \cite{Brotherston14}.
That is, our class neither contains the class of \cite{Brotherston14} nor is contained by it. 

The basic idea of our decision procedure is a novel translation of a given entailment into an equivalent formula in Presburger arithmetic. 
The key idea used in the translation is the notion of ``{\em sorted}'' symbolic heaps. Any heap represented by a sorted symbolic heap has addresses arranged in the order of the spatial part of the symbolic heap. 
If we assume the both sides of given entailment are sorted, 
the entailment is valid if no contradiction is found in comparing spatial parts on both sides starting from left to right. 

We also propose two ideas for improving the performance of our decision procedure. 
The performance heavily depends on the size (number of the separating conjunction symbol $*$) of a given entailment. 
Consider a single conclusion entailment $\phi_1\vdash\phi_2$. Let $n$ and $m$ be numbers of the separating conjunction in $\phi_1$ and $\phi_2$, respectively. 
Then this entailment will be decomposed into $n!$ sorted entailments with $m!$ disjunctions on the right-hand side.
So it is quite important to reduce the number of $*$ as much as possible at an early stage of the procedure. 

This paper also presents our entailment checker $\Checker$, which is an implementation of our decision procedure. 
$\Checker$ first 
(1) optimizes a given entailment according to the improvement idea mentioned above, 
(2) decomposes the resulting entailment into some sorted entailments, then 
(3) translates the decomposed entailments into the corresponding Presburger formulas, and finally 
(4) checks their validity by invoking an external SMT solver $\ZZZ$~\cite{Z3}. The original entailment is answered valid if and only-if all of the decomposed sorted entailments are valid. 
The improvement techniques made our system run in a second in most cases of experiments,
and made our system 200 times faster in some cases. 

We introduce our system of separation logic with arrays in Section 2. 
Section 3 defines the decision procedure of the entailment problem of the system. 
The correctness of the decision procedure is shown in Section 4. 
Two improvement ideas of the decision procedure are discussed in Section 5. 
Section 6 discusses the entailment checker $\Checker$
based on this decision algorithm,
and evaluates its performance with experimental data. 
Section 7 concludes.

\section{Separation Logic with Arrays}

This section defines the syntax and semantics of our separation logic with arrays. 
We first give the separation logic with arrays $\SLG$ in an ordinary style. 
Then we define the symbolic-heap system $\SLAR$ as a fragment of $\SLG$. 

\subsection{Syntax of System of Separation Logic with Arrays}

We have first-order variables $x,y,z,\ldots\in \Vars$ and constants $0,1,2,\ldots$. 
The syntax of $\SLG$ is defined as follows: 

Terms $t ::= x\ |\ 0\ |\ 1\ |\ 2\ | ... |\ t+t$. 

Formulas $\varphi ::= t = t\ |\ \varphi \land \varphi\ |\ \neg \varphi\ |\ \exists x\varphi\ |\ \Emp\ |\ t \mapsto t\ |\ \Array(t,t)\ |\ \varphi * \varphi$.

An atomic formula of the form $t\mapsto u$ or $\Array(t,u)$ is called 
a points-to atomic formula or an array atomic formula, respectively. 
The truth of each formula is interpreted under a state of variables and a heap: 
$\Emp$ is true when the heap is empty; 
$t \mapsto u$ is true when
the heap only has a single memory cell of address $t$ that contains the value $u$; 
$\Array(t,u)$ is true when the heap only has an array 
of index from $t$ to $u$; 
a separating conjunction $\varphi_1 * \varphi_2$ is true when
the heap is split into two disjoint sub-heaps, 
$\varphi_1$ is true under one, and $\varphi_2$ is true under the other. 
The formal definition of these interpretation is given in the next subsection.

The set of free variables (denoted by $\FV(\varphi)$) of $\varphi$ is defined as usual. We also define $\FV(\Vec{\varphi})$ as the union of $\FV(\varphi)$, where $\varphi \in \Vec{\varphi}$. 

We sometimes use the symbol $\sigma$ to denote $\Emp$, $t \mapsto u$, or $\Array(t,u)$. 

We use abbreviations $\varphi_1 \vee \varphi_2$, $\varphi_1 \to \varphi_2$, and $\forall x\varphi$ 
defined in a usual way.
We also write $t \neq u$, $t \le u$, $t < u$, and $\True$ 
as abbreviations of 
$\neg (t = u)$, $\exists x.(u = t + x)$, $t + 1 \le u$, 
and $0 = 0$, respectively. 

A formula is said to be \textit{pure} if it 
is a formula of Presburger arithmetic.

\subsection{Semantics of System of Separation Logic with Arrays}

Let $N$ be the set of natural numbers. We define the following semantic domains: 

$\Val \eqDef N$, 
\quad
$\Loc \eqDef N\setminus\{0\}$, 
\quad
$\Stores \eqDef \Vars \to \Val$, 
\quad
$\Heaps \eqDef \Loc \to_{\rm fin} \Val$. 

$\Loc$ means addresses of heaps. $0$ means Null. 
An element $s$ in $\Stores$ is called a \textit{store}
that means a valuation of variables. 
An element $h$ in $\Heap$ is called a \textit{heap}. 
The domain of $h$ (denoted by $\Dom(h)$) means the memory addresses which are currently used. 
$h(n)$ means the value at the address $n$ if it is defined.
We sometimes use notation $h_1+h_2$ for the disjoint union of $h_1$ and $h_2$,
that is, it is defined when $\Dom(h_1)$ and $\Dom(h_2)$ are disjoint sets, and 
$(h_1+h_2)(n)$ is $h_i(n)$ if $n \in \Dom(h_i)$ for $i = 1,2$.
A pair $(s,h)$ is called a \textit{heap model}. 

The interpretation $s(t)$ of a term $t$ by $s$ is defined by 
extending the definition of $s$ by $s(n) = n$ for each constant $n$, 
and $s(t+u) = s(t)+s(u)$. 

The interpretation $s,h\models \varphi$ of $\varphi$ under the heap model $(s,h)$ 
is defined inductively as follows: 

$s,h\models t = u$
\ iff\ 
$s(t) = s(u)$, 

$s,h\models\varphi_1\land\varphi_2$
\ iff\ 
$s,h\models\varphi_1$ and $s,h\models\varphi_2$, 

$s,h\models \neg\varphi$
\ iff\ 
$s,h\not\models\varphi$, 

$s,h\models \exists x\varphi$
\ iff\ 
$s[x:= a],h\models\varphi$ for some $a \in \Val$, 

$s,h\models \Emp$
\ iff\ 
$\Dom(h) = \emptyset$, 

$s,h\models t \mapsto u$
\ iff\ 
$\Dom(h) = \{s(t)\}$ and $h(s(t)) = s(u)$, 

$s,h\models \Array(t,u)$
\ iff\ 
$\Dom(h) = \{ x \in N\ |\ s(t) \le x \le s(u)\}$ and $s(t) \le s(u)$, 

$s,h\models \varphi_1*\varphi_2$
\ iff\ 
$s,h_1\models \varphi_1$, $s,h_2\models \varphi_2$, and $h = h_1+h_2$ for some $h_1$, $h_2$. 

We sometimes write $s\models \varphi$ if $s,h\models \varphi$ holds for any $h$. 
This notation is mainly used for pure formulas, 
since their interpretation do not depend on the heap-part of heap models. 
We also write $\models \varphi$ if $s,h\models \varphi$ holds for any $s$ and $h$.

The notation $\varphi \models \psi$ is an abbreviation of $\models \varphi \to \psi$, that is, $s,h\models\varphi$ implies $s,h\models\psi$ for any $s$ and $h$. 

Let $I$ be a finite set.
In this paper we implicitly assume a linear order on $I$ 
(this order will be used in the definition of the translation $P$ given in the next section). 
Then we sometimes write $\{\phi_i \mid i \in I\}$ for
$\bigvee\{\phi_i \mid i \in I\}$, that is,
disjunction $\bigvee_{i\in I}\phi_i$ of formulas $\phi_i$ ($i\in I$) under the order of $I$. 
We sometimes abbreviate $\{\phi_i\ |\ i \in I\}$ by $\{\phi_i\}_{i \in I}$. 
It may be further abbreviated by $\Vec\phi$ when $I$ is not important. 

\subsection{Symbolic-Heap System with Arrays}

The symbolic-heap system $\SLAR$ is defined as a fragment of $\SLG$. 
The syntax of $\SLAR$ is given as follows. 
Terms of $\SLAR$ are the same as the terms of $\SLG$. 
Formulas of $\SLAR$ (called \textit{symbolic heaps}) have the following form:

$\phi ::= \exists \Vec x(\Pi \land \Sigma)$

\noindent where $\Pi$ is a pure formula of $\SLG$ 
and $\Sigma$ is the spatial part defined by 

$\Sigma ::= \Emp\ |\ t \mapsto t\ |\ \Array(t,t)\ |\ \Sigma * \Sigma$.

We sometimes write $\exists\Vec x\Sigma$ as an abbreviation of $\exists\Vec x(\True \land \Sigma)$. 
We use notations $\Pi_\phi$ and $\Sigma_\phi$ that mean the pure part
and the spatial part of $\phi$, respectively.

In this paper, we consider \textit{entailments} of $\SLAR$ that have the form: 

$\phi \prove \{\phi_i\ |\ i \in I\}$\qquad ($I$ is a finite set)

The symbolic heap on the left-hand side is called the antecedent of the entailment.
The symbolic heaps on the right-hand side are called the succedents of the entailment.
As we noted before, the right-hand side $\{\phi_i\ |\ i \in I\}$ of an entailment means the disjunction of the symbolic heaps $\phi_i$ ($i\in I$). 

An entailment $\phi \prove \{\phi_i\ |\ i \in I\}$ is said to be \textit{valid}
if
$\phi \models \{\phi_i \ |\ i \in I\}$ holds. 

A formula of the form $\Pi\land\Sigma$ is called a \textit{QF symbolic heap} (denoted by $\varphi$).
Note that existential quantifiers may appear in the pure part of a QF symbolic heap. 
We can easily see that
$\exists\Vec{x}\varphi\models\Vec\phi$ is equivalent to
$\varphi\models \Vec\phi$.
So we often assume that the left-hand sides of entailments are QF symbolic heaps. 

 We call entailments of the form $\varphi \prove \{\varphi_i\ |\ i \in I\}$ \textit{QF entailments}. 

\subsection{Analysis/Verification of Memory Safety}

We intend to use our entailment checker for a part of our
analysis/verification system for memory safety.
We briefly explain it for motivating our entailment checker.

The target programming language is essentially the same as
that in \cite{Reynolds02} except we extend the allocation command for
allocating more than one cells.
We define our programming language in programming language C style.

Expressions $e ::= x \ |\  0 \ |\  1 \ |\  2 \ldots \ |\ e+e$.

Boolean expressions $b ::= e==e \ |\ e<e \ |\ b \&\& b \ |\ b \| b \ |\ !b$.

Programs $P ::= x=e; \ |\ \If (b) \{ P \} \Else \{ P \}; \ |\
\While (b) \{ P \}; \ |\ 
P\ P \ |\ $

\qquad \qquad
$x=\Malloc(y); \ |\
x=*y; \ |\
*x=y; \ |\
\Free(x);$.

$x=\Malloc(y);$ allocates $y$ cells and set $x$ to
the pointer to the first cell. Note that this operation may fail
if there is not enough free memory.

Our assertion language is a disjunction of symbolic heaps, namely,

Assertions $A ::= \phi_1 \lor \dots \lor \phi_n$.

In the same way as \cite{Reynolds02},
we use a triple $\Tri A P B$ that means that
if the assertion $A$ holds at the initial state and the program $P$
is executed, then (1) if $P$ terminates then the assertion $B$ holds
at the resulting state, and (2) $P$ does not cause any memory errors.

As inference rules for triples,
we have ordinary inference rules for Hoare triples including the consequence
rule, as well as the following rules for memory operations.
We write $\Arr2(x,y)$ for $\exists z(\Arr(x,z) \land x+y=z+1)$.
$\Arr2(x,y)$ means the memory block at address $x$ of size $y$.
We sometimes write a formula that is not a disjunction of symbolic heaps
for an equivalent assertion obtained
by ordinary logical equivalence rules.
We also write $x \mapsto \Underscore$ for
$\exists z(x \mapsto z)$ where $z$ is fresh.
\[
\Tri A{x=\Malloc(y);}{\exists x'(A[x:=x'] \land (x=\Nil \lor \Arr2(x,y[x:=x']))}, \\
\Tri{A * y \mapsto t}{x=*y;}{\exists x'(A[x:=x'] * y \mapsto t[x:=x'] \land
x=t[x:=x'])} \ (x'\ \rm fresh), \\
\Tri{A * x \mapsto \Underscore)}{*x=y;}{A * x \mapsto y}, \\
\Tri{A * x \mapsto \Underscore)}{\Free(x);}{A}.
\]

In order to prove memory safety of a program $P$ under a precondition $A$,
it is sufficient to
show that
$\{ A \} P \{ \True \}$ is provable.

By separation logic with arrays, we can show a triple
$\Tri A{x=\Malloc(y);}{A \land x=\Nil \lor
A * \Arr2(x,y) \land x \ne \Nil}$,
but
it is impossible without arrays since $y$ in $\Malloc(y)$ is a variable.
With separation logic with arrays,
we can also show
$\Tri{\Arr(p,p+3)}{*p = 5;}{p \mapsto 5 * \Arr(p+1,p+3)}$.

For the consequence rule
\[
\infer{\Tri A P B}{\Tri {A'} P {B'}} \qquad ({\rm if\ } A \imp A', B' \imp B)
\]
we have to check the side condition $A \imp A'$.
Let $A$ be $\phi_1 \lor \ldots \lor \phi_n$ and
$A'$ be $\phi'_1 \lor \ldots \lor \phi'_m$.
Then we will use our entailment checker to decide
$\phi_i \prove \phi'_1, \ldots, \phi'_m$ for all $1 \le i \le n$.

\subsection{Other Systems of Symbolic Heaps with Arrays}

We will impose the following restrictions for correctness of
our entailment decision procedure:
For a given entailment,
if any array atomic formula in the succedent has the form $\Array(t,t+u)$
such that the term $u$ does not contain any existential variables.

Other known systems of symbolic heaps with arrays are only
the system given in Brotherston et al.~\cite{Brotherston17}.
They gave an independent condition for decidability of the entailment problem of the same symbolic-heap system. 
Their condition disallows existential variables in $u$
for each points-to predicate $t\mapsto u$ in the succedent of an entailment.
In order to clarify the difference between our condition and their condition,
we consider the following entailments:

(i)
$\Array(x,x) \vdash x \mapsto 0, \exists y(y > 0 \land x \mapsto y)$

(ii)
$\Array(1,5) \vdash \exists y,y'(\Array(y,y+1) * \Array(y',y'+2))$

(iii)
$\Array(1,5) \vdash \exists y(\Array(1,1+y) * \Array(2+y,5))$

(iv)
$\Array(1,5) \vdash \exists y,y'(\Array(1,1+y) * 2+y \mapsto y' * \Array(3+y,5))$

The entailment (i) can be decided by our decision procedure, but it cannot be decided by their procedure. 
The entailment (ii) is decided by both theirs and ours. 
The entailment (iii) is decided by theirs, but it does not satisfy our condition. 
The entailment (iv) is decided by neither theirs nor ours. 

Our system and the system in \cite{Brotherston17} have the
same purpose, namely, analysis/verification of memory safety.
Basically their target programming language and assertion language are
the same as ours given in this section.
These entailment checkers are essentially used for deciding the side condition
of the consequence rule.
As explained above, ours and theirs have different restrictions for decidability.
Hence the class of programs is the same for
ours and theirs, but some triples can be proved only by ours and
other triples can be proved only by theirs,
according to the shape of assertions.
We explain them by example.

The triple
\\
$\Tri{p \mapsto \Underscore}{y=x+2; *p=y; x=2; y=3;}
{\exists x'y'(y'=x'+1 \land  p \mapsto y')}$
\\
is true and provable in our system, but it is not provable in their system,
since it is proved by the assignment statement rule and
the consequence rule with
the side condition
$\exists x'y'y''(y'=x'+2 \land x=2 \land y=3 \land p \mapsto y') \prove
\exists x'y'(y'=x'+1 \land p \mapsto y')$,
and
this entailment is out of their system.
On the other hand,
the following triple, 
which is slightly different from but very similar to it,
\\
$\Tri{p \mapsto \Underscore}{y=1; *p=y; x=2; y=3;}
{\exists y'(y'=1 \land p \mapsto y')}$
\\
is true and provable in both of our and their systems.

The triple
$\{\Emp\}\hskip 0.5ex{\tt y=x+2; p=\Malloc(y); x=2; y=3;}$
\\ \hbox{ }\qquad
$\{\exists x'y'(y'=x'+1 \land (p=\Nil \lor \Arr2(p,y')))\}$
\\
is true and provable in their system, but it is not provable in our system,
since it is proved by the assignment statement rule and
the consequence rule with
the side condition
$\exists x'y'y''p'(y'=x'+2 \land x=2 \land y=3 \land (p=\Nil \lor \Arr2(p,y')))
\prove
\exists x'y'(y'=x'+1 \land (p=\Nil \lor \Arr2(p,y')))$,
and
this entailment is out of our system.
On the other hand,
the following triple, 
which is slightly different from but very similar to it,
\\
$\Tri{\Emp}{y=1; p=\Malloc(y); x=2; y=3;}
{\exists x'y'(y'=1 \land (p=\Nil \lor \Arr2(p,y')))}$
\\
is true and provable in both of our and their systems.

All these kinds of triples are necessary for program verification in the
real world, and in this sense both our system and their system
have advantage and disadvantage.
We can use both systems together to prove a single triple,
by using one of them to check each necessary side condition,
depending on the shape of the side condition.

Note that 
our and their restrictions
are not determined by a program syntax and they
depend on a context of an assignment statement.
 It is because both our and their restrictions are involved in
existential variables,
which are
$\exists x'$
introduced by the assignment statement \verb|x:=e;|,
and a program syntax cannot decide
where $x'$ appears in a current assertion.

\section{Decision Procedure}\label{sec:decidability}

\subsection{Sorted Entailments}

This subsection describes our key idea, namely {\em sorted} symbolic heaps.
The addresses of heaps represented by a sorted symbolic heap must be sorted, that is,
their order can be determined by the order of the spatial part of the sorted symbolic heap. 

We sometimes regard $\Sigma$ as a list of $\Emp$, $t\mapsto u$, and $\Array(t,u)$. 
We also regard the symbol $*$ written like $t\mapsto u * \Sigma$ as the list constructor. 
By abuse of notation, we write $\Emp$ in order to represent the empty list. 

A symbolic heap $\phi$ is called {\em sorted} at $(s,h)$ 
if $s,h$ satisfies $\phi$ and 
the addresses of the heap $h$ are arranged in the order of the spatial part of $\phi$.

In order to express this notion, we introduce pure formulas $t < \Sigma$ and $\Sorted(\Sigma)$, 
which mean the first address expressed by $\Sigma$ is greater than $t$, 
and $\Sigma$ is sorted, respectively. 
They are inductively defined as follows: 

$t < \Emp \eqDef \True$, 
\qquad
$t < (\Emp * \Sigma_1) \eqDef t < \Sigma_1$, 

$t < (t_1 \mapsto u_1 * \Sigma_1) \eqDef t < t_1$, 
\qquad
$t < (\Array(t_1,u_1) * \Sigma_1) \eqDef t < t_1$, 

$\Sorted'(\Emp) \eqDef \True$, 

$\Sorted'(\Emp*\Sigma_1) \eqDef \Sorted'(\Sigma_1)$, 

$\Sorted'(t\mapsto u * \Sigma_1) \eqDef t < \Sigma_1 \land \Sorted'(\Sigma_1)$, 

$\Sorted'(\Array(t,u) * \Sigma_1) \eqDef t \le u \land u < \Sigma_1 \land \Sorted'(\Sigma_1)$,

$\Sorted(\Sigma) \eqDef 0 < \Sigma \land \Sorted'(\Sigma)$. 

The formula $\Sorted(\Sigma) \land \Sigma$ is sometimes 
abbreviated by $\Tilde \Sigma$ or $\Sigma^\sim$.
We also write $\Tilde \phi$ (or $\phi^\sim$) for 
the symbolic heap which is obtained from $\phi$ by replacing 
$\Pi_\phi$ by $\Pi_\phi\land\Sorted(\Sigma_\phi)$. 

We claim that
$\phi$ is a sorted symbolic heap at $(s,h)$ 
iff $\Tilde\phi$ is true in $(s,h)$. 

We define $\Perm(\Sigma)$ as the set of permutations of $\Sigma$ with respect to $*$.
A symbolic heap $\phi'$ is called a permutation of $\phi$
if $\Sigma_{\phi'} \in \Perm(\Sigma_\phi)$ and the other parts of $\phi$ and $\phi'$ are same. 
We write $\Perm(\phi)$ for the set of permutations of $\phi$. 

Note that $s,h\models\phi$ iff $s,h\models\Tilde{\phi'}$ for some $\phi'\in\Perm(\phi)$. 

An entailment is said to be sorted if all of its antecedent and succedents have the form $\Tilde\phi$. 
We claim that checking validity of entailments can be reduced to checking validity of sorted entailments. 
The formal statement of this property will be given later (see Lemma~\ref{lemma:split_sorted}). 

The basic idea of our decision procedure is as follows:
(1) A given entailment is decomposed into sorted entailments according to Lemma~\ref{lemma:split_sorted}; 
(2) the decomposed sorted entailments are translated into Presburger formulas by the translation $P$ given in the next subsection; 
(3) the translated formulas are decided by the decision procedure of Presburger arithmetic. 

\subsection{Translation $P$}
We define the translation $P$ from QF entailments into Presburger formulas. 
We note that the resulting formula may contain new fresh variables (denoted by $z$).

For saving space, we use some auxiliary notations.
Let $\{t_j\}_{j\in J}$ be a set of terms indexed by a finite set $J$. 
We write $u = t_J$ for $\bigwedge_{j\in J} u = t_j$.
We also write $u < t_J$ for $\bigwedge_{j\in J} u < t_j$.

The definition of $P(\Pi,\Sigma,S)$ is given as listed in Fig.~\ref{fig:transP}, where $S$ is a finite set $\{(\Pi_i,\Sigma_i)\}_{i \in I}$. 
We assume that pattern-matching is done from top to bottom. 

In order to describe the procedure $P$,
we temporarily extend terms to include  $u - t$ 
where $u,t$ are terms.
In the result of $P$, which is a 
Presburger arithmetic formula,
we eliminate these extended terms by replacing
$t'+(u-t)=t''$ and $t'+(u-t)<t''$
by
$t'+u = t''+ t$ and $t'+u < t'' + t$, respectively.

The $\eqDef$ steps terminate
since
$(|\Sigma|+\sum_{i \in I}|\Sigma_i|, |S|)$ decreases
where
$|\Sigma|$ is the number of $*$ in $\Sigma$
and $|S|$ is the number of elements in $S$.
Note that this measure does not decrease for some $\eqDef$, but
the left-hand sides of the definition are mutually exclusive and
hence combination of $\eqDef$ eventually decreases the measure.
For example, ($\mapsto\mapsto$) will eventually come after ({\bf Arr}$\mapsto$). 

\begin{figure}[t]\small
  \rule{\textwidth}{1pt}
  \\[10pt]
$
\begin{array}{lll}
  P(\Pi,\Emp * \Sigma,S)
  &\eqDef&
  P(\Pi,\Sigma,S)
  \hspace{120pt}
  {\bf (EmpL)}
  \\
  P(\Pi,\Sigma,\{( \Pi',\Emp * \Sigma')\} \cup S)
  &\eqDef&
  P(\Pi,\Sigma,\{( \Pi',\Sigma')\} \cup S)
  \hfill
  {\bf (EmpR)}
  \\
  P(\Pi,\Emp,\{( \Pi',\Sigma')\} \cup S)
  &\eqDef&
  P(\Pi,\Emp,S),
  \quad \hbox{where $\Sigma' \not\equiv \Emp$}
  \hfill
  {\bf (EmpNEmp)}
  \\
  P(\Pi,\Emp,\{( \Pi_i,\Emp)\}_{i \in I})
  &\eqDef&
  \Pi \imp \bigvee_{i \in I} \Pi_i
  \hfill
  {\bf (EmpEmp)}  
  \\
  P(\Pi,\Sigma,\{( \Pi', \Emp)\} \cup S)
  &\eqDef&
  P(\Pi,\Sigma,S), 
  \qquad\hbox{where $\Sigma \not\equiv \Emp$}
  \hfill
  {\bf (NEmpEmp)}  
  \\
  P(\Pi,\Sigma,\emptyset)
  &\eqDef&
  \neg(\Pi \land \Sorted(\Sigma))
  \hfill
  {\bf (empty)}    
\end{array}
$
\\[5pt]
$P(\Pi,t \mapsto u * \Sigma,\{( \Pi_i, t_i \mapsto u_i * \Sigma_i)\}_{i \in I})$
\hfill{($\mapsto\mapsto$)}
\\
\begin{tabular}[t]{ll}
  $\eqDef$&
  $P(\Pi \land t < \Sigma,\Sigma, \{( \Pi_i \land t=t_i \land u=u_i \land t_i < \Sigma_i,\Sigma_i)\}_{i \in I})$
\end{tabular}  
\\[5pt]
$P(\Pi,t \mapsto u * \Sigma,\{( \Pi_i,\Array(t_i,t_i') * \Sigma_i)\} \cup S)$
\hfill{\bf ($\mapsto$Arr)}
\\
\begin{tabular}[t]{ll}
  $\eqDef$&
  $P(\Pi \land t_i'=t_i,t \mapsto u * \Sigma,\{(\Pi_i,t_i \mapsto u * \Sigma_i)\} \cup S)$
  \\
  &
  $\land\
  P(\Pi \land t_i'>t_i,t \mapsto u * \Sigma,\{(\Pi_i,t_i \mapsto u * \Array(t_i+1,t_i') * \Sigma_i)\} \cup S)$
  \\
  &
  $\land\
  P(\Pi \land t_i'<t_i,t \mapsto u * \Sigma,S)$
\end{tabular}
\\[5pt]
$P(\Pi,\Array(t,t') * \Sigma,S)$
\hfill{\bf (Arr$\mapsto$)}
\\
\begin{tabular}[t]{ll}
  $\eqDef$&
  $P(\Pi \land t' > t,t \mapsto z * \Array(t+1,t') * \Sigma,S)$
  \\
  &
  $\land\ P(\Pi \land t' = t,t \mapsto z' * \Sigma,S)$,
  \quad
  \hbox{where $(\Pi'', t'' \mapsto u'' * \Sigma'') \in S$ and $z,z'$ are fresh}
\end{tabular}
\\[5pt]
$P(\Pi,\Array(t,t') * \Sigma,\{( \Pi_i,\Array(t_i,t_i') * \Sigma_i)\}_{i \in I})$
\hfill{\bf (ArrArr)}
\\
\begin{tabular}[t]{ll}
  $\eqDef$&
  $\Land_{I' \subseteq I}P\left(
  \begin{array}{l}
    \Pi \land
    m = m_{I'} \land m < m_{I\setminus I'}
    \land t \le t' \land t' < \Sigma, \Sigma,
    \\
    \{( \Pi_i \land t_i+m < \Sigma_i, \Sigma_i)\}_{i \in I'}
    \cup
    \{( \Pi_i,\Array(t_i+m+1,t_i') * \Sigma_i)\}_{i \in I\setminus I'}
  \end{array}
  \right)$
  \\
  \multicolumn{2}{l}{
  $\land
  \Land_{\emptyset \ne I' \subseteq I}
  P\left(
  \begin{array}{l}
    \Pi
    \land m'<m \land m' = m_{I'} \land  m'<m_{I\setminus I'},
    \Array(t+m'+1,t') * \Sigma,
    \\
    \{( \Pi_i \land t_i+m' < \Sigma_i, \Sigma_i )\}_{i \in I'}
    \cup
    \{( \Pi_i,\Array(t_i+m'+1,t_i') * \Sigma_i )\}_{i \in I\setminus I'}
  \end{array}
  \right)$, 
  }
\end{tabular}
\\
where $m$, $m_i$, and $m'$ are abbreviations of $t'-t$, $t_i'-t_i$, and $m_{\min I'}$, respectively. 
\rule{\textwidth}{1pt}
\caption{The translation $P$}
\label{fig:transP}
\end{figure}

The formula $P(\Pi,\Sigma,\{(\Pi_i,\Sigma_i)\}_{i\in I})$ means 
that the QF entailment 
$\Pi\land\Tilde\Sigma \vdash \{\Pi_i\land\Tilde\Sigma\}_{i \in I}$ is valid 
(in fact, we can show their equivalence by induction on the definition of $P$).
From this intuition, we sometimes call $\Sigma$ the left spatial formula, and also call $\{\Sigma_i\}_{i\in I}$ the right spatial formulas.
We call the left-most position of a spatial formula
the head position.
The atomic formula appears at the head position is called the head atom. 

We will explain the meaning of each clause in Fig.\ref{fig:transP}. 

The clauses {\bf (EmpL)} and {\bf (EmpR)} just remove $\Emp$
at the head position. 

The clause {\bf (EmpNEmp)} handles the case where
the left spatial formula is $\Emp$.
A pair $(\Pi',\Sigma')$ in the third argument of $P$ is removed
if $\Sigma'$ is not $\Emp$,
since $\Pi'\land\Sigma'$ cannot be satisfied by the empty heap. 

The clause {\bf (EmpEmp)} handles the case where 
the left formula and all the right spatial formulas are $\Emp$.
This case $P$ returns a Presburger formula
which is equivalent to the corresponding entailment is valid. 

The clause {\bf (NEmpEmp)} handles the case where
the left spatial formula is not $\Emp$ and
a pair $(\Pi',\Emp)$ appears in the third argument of $P$. 
We remove the pair since $\Pi'\land\Emp$ cannot be satisfied
by any non-empty heap. 
For example,
$P(\True,x\mapsto 0 * y\mapsto 0,\{(\True,\Emp)\})$
becomes 
$P(\True,x\mapsto 0 * y\mapsto 0,\emptyset)$. 

The clause {\bf (empty)} handles the case where
the third argument of $P$ is empty. 
This case $P$ returns a Presburger formula which is equivalent to 
that the left symbolic heap $\Pi\land\Sigma$ is not satisfiable. 
For example,
$P(\True,x\mapsto 0 * y\mapsto 0,\emptyset)$
returns 
$\neg(\True \land x < y)$. 

The clause {\bf ($\mapsto\mapsto$)} handles the case where
all the head atoms of $\Sigma$ and $\{\Sigma_i\}_{i\in I}$ are
the points-to predicate. 
This case we remove all of them and put equalities on the right pure parts. 
By this rule the measure is strictly reduced. 
For example,
$P(3<4,3\mapsto 10 * 4\mapsto 11,\{(\True,3\mapsto 10 * \Array(4,4))\})$
becomes

$P(3<4\land 3<4,4\mapsto 11,\{(\True\land 3=3 \land 4=4\land 3<4, \Array(4,4))\})$

This can be simplified to
$P(3<4,4\mapsto 11,\{(3=3 \land 4=4\land 3<4, \Array(4,4))\})$,
since
$3<4\land 3<4$ is logically equivalent to $3<4$
and
$\True\land 3=3$ is logically equivalent to $3=3$. 
In the following examples, we implicitly use similar simplifications. 

The clause {\bf ($\mapsto$Arr)} handles the case where
the head atom of the left spatial formula is the points-to predicate and
some right spatial formula $\Sigma_i$ has the array predicate as its head atom.
Then we split the array atomic formula into the points-to and the rest. 
We have three subcases according to the length of the head array. 
The first case is when the length of the array is $1$: 
We replace the head array by a points-to atomic formula. 
The second case is when the length of the head array is
greater than $1$: 
We split the head array into the points-to predicate and the rest array. 
The last case is when the length of the head array is less than $1$: 
We just remove $(\Pi_i,\Sigma_i)$, since the array predicate is false. 
We note that this rule can be applied repeatedly until all head arrays of
the right spatial formulas are unfolded,
since the left spatial formula is unchanged.
Then the measure is eventually reduced by applying {\bf ($\mapsto\mapsto$)}. 
For example,
$P(\True,4\mapsto 11,\{(3=3\land10=10,\Array(4,4))\})$
becomes

$P(3<4\land 4=4, 4\mapsto 11,\{(3=3\land 10=10,4\mapsto 11)\})$

\quad
$\land
P(3<4\land 4<4, 4\mapsto 11,\{(3=3\land 10=10,4\mapsto 11 * \Array(5,4))\})$

\quad
$\land
P(3<4\land 4>4, 4\mapsto 11,\emptyset)$.

The clause {\bf (Arr$\mapsto$)} handles the case where
the head atom of the left spatial formula is array and 
there is a right spatial formula whose head atom is the points-to predicate.
We have two subcases according to the length of the head array. 
The first case is when the length of the array is $1$: 
The array is unfolded and it is replaced by a points-to atomic formula with
a fresh variable $z$. 
The second case is the case where the length of the array is greater than $1$: 
The array is split into the points-to predicate (with a fresh variable $z'$)
and the rest array.
We note that the left head atom becomes a points-to atomic formula
after applying this rule. 
Hence the measure is eventually reduced,
since {\bf ($\mapsto\mapsto$)} or {\bf ($\mapsto$Arr)} will be applied next. 
For example,
$P(\True,\Array(x,x),\{(\True,x\mapsto 10)\})$
becomes

$P(x>x, x\mapsto z*\Array(x+1,x),\{(\True,x\mapsto 10)\})$

\quad
$\land
P(x=x, x\mapsto z',\{(\True,x\mapsto 10)\})$. 

The last clause {\bf (ArrArr)} handles the case where
all the head atoms in the left and right spatial formulas are arrays. 
We first find the head arrays with the shortest length among the head arrays. 
Next we split each longer array into two arrays
so that the first part has the same size to the shortest array. 
Then we remove the first parts. The shortest arrays are also removed. 
In this operation we have two subcases: 
The first case is when the array of the left spatial formula
has the shortest size and disappears by the operation. 
The second case is when the array of the left spatial formula
has a longer size, it is split into two arrays, and the second part remains.
We note that the measure is strictly reduced,
since at least one shortest array is removed.
For example,
$P(\True,\Array(3,5),\{(\True,\Array(3,3)*\Array(4,5))\})$
becomes

$P(2<0 \land 3\le 5, \Emp,\{(\True,\Array(6,3)*\Array(4,5))\})$

\quad
$\land
P(2=0 \land 3\le 5, \Emp,\{(\True\land 3<4,\Array(4,5))\})$

\quad
$\land
P(0<2\land 0=0, \Array(4,5),\{(\True,\Array(4,5))\})$. 

Note that 
the sizes of $\Arr(3,5)$ and $\Arr(3,3)$ are 3 and 1 respectively,
and we have three cases for 
$2<0$, $2=0$, and $0<2$,
by comparing them (actually comparing (them - 1)).
\\

\noindent{\bf Example.}\quad
The sorted entailment
$(3\mapsto 10*4\mapsto 11)^\sim \vdash \Array(3,4)^\sim$
is translated by computing
$P(\True,3\mapsto 10*4\mapsto 11, \{(\True,\Array(3,4))\})$. 
We will see its calculation step by step. 
It first becomes

$P(3=4,3\mapsto 10 * 4\mapsto 11,\{(\True,3\mapsto 10)\})$

\quad
$\land
P(3<4,3\mapsto 10 * 4\mapsto 11,\{(\True,3\mapsto 10 * \Array(4,4))\})$

\quad
$\land
P(3>4,3\mapsto 10 * 4\mapsto 11,\emptyset)$
\\
by ($\mapsto$Arr). 
The first conjunct becomes 
$P(3=4\land 3<4, 4\mapsto 11,\{(3=3\land 10=10,\Emp)\})$
by $(\mapsto\mapsto)$,
then it becomes $\neg(3=4\land 3<4)$ by (NEmpEmp) and (empty).
The third conjunct becomes $\neg(3>4\land 3<4)$ by (empty). 
The second conjunct becomes 
$P(3<4, 4\mapsto 11,\{(3=3\land 10=10,\Array(4,4))\})$ by $(\mapsto\mapsto)$,
then it becomes

$P(3<4\land 4=4, 4\mapsto 11,\{(3=3\land 10=10,4\mapsto 11)\})$

\quad
$\land
P(3<4\land 4<4, 4\mapsto 11,\{(3=3\land 10=10,4\mapsto 11 * \Array(5,4))\})$

\quad
$\land
P(3<4\land 4>4, 4\mapsto 11,\emptyset)$
\\
by ($\mapsto$Arr). Hence we have

$P(3<4\land 4=4, \Emp,\{(3=3\land 10=10\land 4=4 \land 11=11,\Emp)\})$

\quad
$\land
P(3<4\land 4<4, \Emp,\{(3=3\land 10=10\land 4=4 \land 11=11,\Array(5,4))\})$

\quad
$\land
\neg(3<4\land 4>4)$
\\
by $(\mapsto\mapsto)$ and (empty).
We note that the second one becomes
$P(3<4\land 4<4, \Emp,\emptyset\})$ by (EmpNEmp). 
Thus we obtain

$(3<4\land 4=4) \to (3=3\land 10=10\land 4=4 \land 11=11)$

\quad
$\land
\neg(3<4\land 4<4)$
$\land
\neg(3<4\land 4>4)$
\\
by (EmpEmp) and (empty). 
Finally we obtain 
$\neg(3=4\land 3<4)
  \land
  (3<4\land 4=4) \to (3=3\land 10=10\land 4=4 \land 11=11)
  \land
  \neg(3<4\land 4<4)
  \land
  \neg(3>4\land 3<4)$. 

In our decision procedure the produced Presburger formula will be checked by an external SMT solver.

\subsection{Decidability}

The aim of $P$ is to give an equivalent formula of Presburger arithmetic to a given entailment. The correctness property of $P$ is stated as follows. 

\begin{theorem}[Correctness of Translation $P$]\label{thm:correct}
If any array atomic formula in $\Sigma_i$ has the form $\Array(t,t+u)$
such that the term $u$ does not contain $\Vec{y}$, then 
\[
\Pi\land\Tilde\Sigma \models \{\exists \Vec{y_i}(\Pi_i\land\Tilde\Sigma_i)\}_{i\in I}
\quad
\hbox{ iff }
\quad
\models \forall \Vec z\exists \Vec yP(\Pi,\Sigma,\{(\Pi_i,\Sigma_i)\}_{i\in I})
\]
where $\Vec y$ is a sequence of $\Vec{y_i}$ ($i\in I$), and
$\Vec z$ is $\FV(P(\Pi,\Sigma,\{(\Pi_i,\Sigma_i)\}_{i\in I}))\setminus \FV(\Pi,\Sigma,\{\Pi_i\}_{i\in I},\{\Sigma_i\}_{i\in I})$. 
\end{theorem}
We note that $\Vec z$ are the fresh variables introduced in the unfolding of $P(\Pi,\Sigma,\{(\Pi_i,\Sigma_i)\}_{i\in I})$. 

The proof of this theorem will be given in the next section. 

The correctness property is shown with the condition described in the theorem. 
This condition avoids a complicated situation for $\Vec y$ and $\Vec z$, 
such that some variables in $\Vec y$ depend on $\Vec z$, and some determine $\Vec z$. 
For example, if we consider $\Array(1,5)\vdash \exists y_1y_2(\Array(1,y_1)*y_1+1\mapsto y_2*\Array(y_1+2,5))$, 
we will have $y_1+1\mapsto z$ during the unfolding of 
$P(\True,\Array(1,5),\{(\True,\Array(1,y_1)*y_1+1\mapsto y_2*\Array(y_1+2,5))\})$. 
Then finally we have $z=y_2$ after some logical simplification. 
This fact means that $y_2$ depends on $z$, 
and moreover $z$ is indirectly determined by $y_1$. 
The latter case occurs when sizes of array depend on $\Vec y$. 
We need to exclude this situation. 

Finally we have the decidability result for the entailment problem of $\SLAR$ under the condition from the above theorem and the property of sorted entailments (stated in Lemma~\ref{lemma:split_sorted}). 

\begin{corollary}[Decidability of Validity Checking of Entailments]\label{cor:decidability}
  Validity checking of entailments
  $\Pi\land\Sigma \vdash \{\exists \Vec{y_i}(\Pi_i\land\Sigma_i)\}_{i\in I}$ of $\SLAR$ 
  is decidable, 
if any array atomic formula in $\Sigma_i$ has the form $\Array(t,t+u)$ such that the term $u$ does not contain $\Vec{y_i}$. 
\end{corollary}

\noindent{\bf Example.}\quad
Our motivating example
$\Array(x,x) \vdash x \mapsto 0, \exists y(y > 0 \land x \mapsto y)$
satisfies the condition, and its validity is checked in the following way. 

- It is decomposed into several sorted entailments: in this case, it produces the same entailment. 

- Compute $P(\True,\Array(x,x), S_1)$, 
where $S_1$ is $\{(\True,x \mapsto 0),(y>0, x \mapsto y)\}$. 
It becomes
$P(x<x, x\mapsto z*\Array(x+1,x), S_1) \land P(x=x, x\mapsto z, S_1)$
by (Arr$\mapsto$). Then it becomes 

$P(x<x\land x<x+1, \Array(x+1,x), S_2) \land P(x=x\land x<x+1, \Emp, S_2)$, 
\\
where $S_2$ is 
$\{(x=x\land z=0,\Emp),(y>0 \land x=x\land z=y,\Emp)\}$. 
The former conjunct becomes $P(x=x \land x<x+1,\Array(x+1,x),\emptyset)$ by (NEmpEmp), 
then, by (empty), it becomes $\neg(x=x \land x<x+1 \land x+1\le x)$, which is equivalent to $\neg(x<x+1\land x+1\le x)$.
The latter conjunct becomes $x=x\land x<x+1 \to (x=x \land z=0)\vee(y>0\land x=x\land z=y)$, 
which is equivalent to $x<x+1 \to z=0\vee(y>0\land z=y)$, 

- Check validity of the formula
$\forall xz\exists yP(\True,\Array(x,x), S_1)$, 
which is equivalent to $\forall xz\exists y(\neg(x<x+1\land x+1\le x) \land (z=0\vee (y>0\land z=y)))$. 
Finally the procedure answers ``valid'', since the produced Presburger formula is valid. 

\begin{remark}
  Our result can be extend to the class of entailments which are semantically equivalent to entailments that satisfy the condition. For example, the entailment 
$\Array(1,10) \vdash \Array(1,x) * \Array(x+1,10)$ 
does not satisfy the condition because the occurrences of the array atomic formulas on the succedent are not in the form required by the condition.
However it can be decided, since it is equivalent to the following entailment which satisfy the condition: 

$x = 1+z \land x+w = 9 \land \Array(1,10) \vdash \Array(1,1+z) * \Array(z+2,(z+2)+w)$.

In other words, our procedure can decide an entailment that the lengths of the array predicate on the succedent do not depend on the existential variables. 
\end{remark}

\section{Correctness of Decision Procedure}\label{sec:correctness}

This section shows correctness of our decision procedure. 
We first show the basic property of sorted entailments. 

\begin{lemma}\label{lemma:split_sorted}
$s\models \varphi\to\bigvee_{i\in I}\phi_i$
is equivalent to 
\\
$s\models \Tilde{\varphi'}\to\Lor\{\Tilde{\phi'}\ |\ i\in I, \phi'\in\Perm(\phi_i)\}$
for all $\varphi' \in \Perm(\varphi)$
\end{lemma}
\Proof
We first show the left-to-right part. 
Assume the left-hand side of the claim. 
Fix $\varphi'\in \Perm(\varphi)$ and suppose $s,h \models \Tilde{\varphi'}$.
Then we have $s,h \models \varphi$.
By the assumption, $s,h\models \phi_i$ for some $i\in I$. 
Hence we have $s,h\models\Lor\{\Tilde{\phi'}\ |\ i\in I, \phi'\in\Perm(\phi_i)\}$. 
Next we show the right-to-left part.
Assume the right-hand side and $s,h\models\varphi$.
We have $s,h\models\Tilde{\varphi'}$ for some $\varphi'\in\Perm(\varphi)$. 
By the assumption, $s,h\models\Tilde{\phi'}$ for some $\phi'\in\Perm(\phi_i)$. 
Thus we have $s,h\models\phi_i$ for some $i\in I$. 
$\Box$
\\

This lemma shows that
validity checking problem of a given entailment can be reduced to
that of several sorted entailments.

\subsection{Correctness of Translation}

This subsection shows correctness of the translation $P$. 
The main difficulty for showing correctness is how to handle the new variables (denoted by $z$)
that are introduced during the unfolding $P$. 
In order to do this, we temporarily extend our language with new terms, denoted by $[t]$. 
A term $[t]$ means the value at the address $t$, that is, 
it is interpreted to $h(s(t))$ under $(s,h)$.
We will use this notation instead of $z$, since $z$ must appear in the form $t \mapsto z$  during unfolding $P$, and this $t$ is unique for $z$. 
Notice that both $s$ and $h$ are necessary for interpreting 
a formula of the extended language even 
if it is a pure formula. 

In this extended language, we temporarily introduce a variant $P'$ of $P$
so that we use $[t]$ instead of $z$, 
which is defined in the same way as $P$ except
\begin{align*}
P'(\Pi,\Array(t,t') * \Sigma,S) 
\eqDef
&
P'(\Pi \land t'=t, t \mapsto [t] * \Sigma,S)
\\
&\land
P'(\Pi \land t'>t, t \mapsto [t] * \Array(t+1,t') * \Sigma,S), 
\end{align*}
when $(\Pi'', t'' \mapsto u'' * \Sigma'') \in S$.
Note that $P'$ never introduces any new variables.

We will introduce some notations. 
Let $S$ be $\{(\Pi,\Sigma)\}_{i\in I}$. 
Then we write $\Tilde S$ for $\{ \Pi_i \land \Tilde{\Sigma_i} \}_{i\in I}$.
We write $\Dom(s,\Sigma)$ for 
the set of addresses used by $\Sigma$ under $s$, that is, 
it is inductively defined as follows: 
$\Dom(s,\Emp) = \emptyset$, 
$\Dom(s,\Emp * \Sigma_1) = \Dom(s,\Sigma_1)$, 
$\Dom(s,t\mapsto u*\Sigma_1) = \{s(t)\}\cup\Dom(s,\Sigma_1)$, and 
$\Dom(s,\Array(t,u)*\Sigma_1) = \{s(t),\ldots,s(u)\}\cup\Dom(s,\Sigma_1)$ if $s(t) \le s(u)$. 

The next lemma clarifies the connections between entailments, $P$, and $P'$. 

\begin{lemma}\label{lemma:correct}
{\rm (1)}
Assume $s,h \models \hat \Pi \land \hat \Sigma$.
Suppose $P'(\Pi,\Sigma,S)$ appears in the unfolding of $P'(\hat\Pi, \hat\Sigma, \hat S)$.
Then

$
s,h|_{\Dom(s,\Sigma)} \models P'(\Pi,\Sigma,S)
$
iff
$
s,h|_{\Dom(s,\Sigma)} \models \Pi \land \Sorted(\Sigma) \imp \Lor\Tilde S.
$

{\rm (2)}
$
\forall sh(s,h \models \Pi\land\Tilde\Sigma \imp s,h \models \exists \Vec yP'(\Pi,\Sigma,S))
$
iff
$
\Pi\land\Tilde\Sigma \models \exists \Vec y\Lor\Tilde S.
$

{\rm (3)} $\models \neg(\Pi \land \Sorted(\Sigma)) \imp P(\Pi,\Sigma,S)$.

{\rm (4)}
$
\forall sh(s,h \models \Pi \land \Tilde\Sigma \imp s,h \models 
\forall \Vec z\exists \Vec yP(\Pi,\Sigma,S))
$
iff
$
\models \forall \Vec z\exists \Vec yP(\Pi,\Sigma,S).
$
\end{lemma}
\noindent\textbf{Proof}. 
The claim (1) is shown by induction on the steps $\eqDef$.
The claim (2) can be obtained by using (1). 
The claim (3) is proved by induction on the steps of $\eqDef$.
The claim (4) is shown by using (3).
$\Box$
\\

Recall that our condition requires that
lengths of arrays on the succedent does not depend on existential variables.
We note that, under our condition, each $t$ that appears as $t \mapsto u$ or $\Array(t,t')$ in the second argument of $P'$ during the unfolding of $P'$ does not contain any existential variables. 
By this fact, we can see that each term $[t]$ does not contain existential variables,
since it first appears as $t \mapsto [t]$ in the second argument of $P'$ during the unfolding of $P'$. 
\\

\noindent\textbf{Proof of Theorem~\ref{thm:correct}}
\quad
Let $S$ be $\{(\Pi_i,\Sigma_i)\}_{i\in I}$.
Then the left-hand side is equivalent to
$\Pi \land \Tilde\Sigma \models \exists\Vec y\Lor \Tilde S$.
Moreover, by Lemma~\ref{lemma:correct}~(2), it is equivalent to
\begin{equation}\label{eq_Z}
\forall sh(s,h \models \Pi \land \Tilde\Sigma \imp s,h \models \exists \Vec yP'(\Pi,\Sigma,S))
\end{equation}
By Lemma~\ref{lemma:correct}~(4), the right-hand side is equivalent to 
\begin{equation}\label{eq_A}
\forall sh\forall \Vec z(s,h \models \Pi \land \Tilde\Sigma \imp s \models \exists \Vec yP(\Pi,\Sigma,S)). 
\end{equation}
Now we will show the equivalence of (\ref{eq_Z}) and (\ref{eq_A}). 
Here we assume $[t_1],\ldots,[t_n]$ appear in $P'(\Pi,\Sigma,S)$
and $s\models t_1 < \ldots < t_n$,
we let $\Vec z = z_1,\ldots,z_n$.
Notice that each $[t_j]$ does not contain any existential variable because of the condition. 
So we can obtain $P'(\Pi,\Sigma,S) = P(\Pi,\Sigma,S)[\Vec z:=[\Vec t]]$. 
Hence (\ref{eq_Z}) is obtained from (\ref{eq_A}) by taking $z_i$ to be $[t_i]$ for $1 \le i \le n$.

We show the inverse direction. Assume (\ref{eq_Z}). 
Fix $s$, $h$, and $\Vec a$ for $\Vec z$.
Let $s'$ be $s[\Vec z:=\Vec a]$.
Let $h'$ be $h[s(\Vec t):=\Vec a]$.
Then by (\ref{eq_Z}), we have 
\[
s,h' \models \Pi \land \Tilde\Sigma \imp s,h' \models \exists \Vec yP'(\Pi,\Sigma,S).
\]
We claim that $s,h'\models \exists\Vec y.P'(\Pi,\Sigma,S)$ is equivalent to
$s'\models \exists\Vec y.P(\Pi,\Sigma,S)$. 
We also claim that 
$s,h' \models \Pi \land \Tilde\Sigma$ is equivalent to
$s,h \models \Pi \land \Tilde\Sigma$, since each $t_i$ appears as an address of an array predicate in $\Sigma$. Therefore we have (\ref{eq_A}). 
$\Box$
\\

\section{Improvement of Decision Procedure}

Our decision procedure is not efficient because of the decomposition. 
The given unsorted entailment is decomposed into some sorted entailments with very large number of succedents.  
Recall that $|\Sigma|$ is the number of $*$ in $\Sigma$. 
Then an unsorted entailment 
$\Pi\land\Sigma\vdash \{\exists\Vec{y_i}(\Pi_i\land\Sigma_i)\}_{i \in I}$ 
is decomposed into $|\Sigma|!$~-sorted entailments with 
$\sum_{i\in I} |\Sigma_i|!$~-succedents. 

We currently adapt the following two ideas to improve this situation. 

\noindent\textbf{(U) Elimination of redundancy by using unsatisfiability checking}\quad
We can easily observe that sorted entailments after decomposition often contain many redundant parts. 
For example,
let $\sigma_1$, $\sigma_2$ and $\sigma_3$ be
$1\mapsto 10$, $2\mapsto 20$ and $3\mapsto 30$, respectively. 
An unsorted entailment 
$\sigma_2 * \sigma_1 \vdash \sigma_1 * \sigma_2, \sigma_3$
is decomposed into the following sorted entailments (after some simplification)
\begin{center}
$2 < 1 \land \sigma_2 * \sigma_1 \vdash 1<2\land\sigma_1 * \sigma_2,\,2<1\land\sigma_2 * \sigma_1,\,\sigma_3$
  \\
$1 < 2 \land \sigma_1 * \sigma_2 \vdash 1<2\land\sigma_1 * \sigma_2,\,2<1\land\sigma_2 * \sigma_1,\,\sigma_3$
\end{center}
The first entailment is trivially valid, since its antecedent is unsatisfiable.
So we can skip checking this entailment. 
The last two succedents of the second entailment are redundant, 
since they never satisfied with the antecedent. 
So we can drop them. 
More formally, we apply the following reduction rule:

$\varphi \vdash \Vec{\phi_1},\phi,\Vec{\phi_2}$ 
is reduced to 
$\varphi \vdash \Vec{\phi_1},\Vec{\phi_2}$ 
if $\varphi\land\phi$ is unsatisfiable. 

\noindent\textbf{(F) Frame elimination by using the invertible frame rule}\quad
Our second improvement is to reduce the number of separating conjunctions in the given unsorted entailment (before the decomposition process). 
This improvement is effective since reducing the number of separating conjunctions reduces the number of the succedents after decomposition. 

In order to reduce separating conjunctions, we use the frame rule: 
$\Pi\land\Sigma \vdash \Pi'\land\Sigma'$ implies 
$\Pi\land\sigma * \Sigma \vdash \Pi'\land\sigma * \Sigma'$, 
where $\sigma$ is $t\mapsto u$ or $\Array(t,u)$. 
However the frame rule of this form is not invertible, that is, the inverse direction of the implication does not generally hold. 
In our setting, since we have inequalities, 
we can define $\Disj(\sigma,\Sigma)$ as a pure formula, 
which means the memory cells used by $\sigma$ and $\Sigma$ are disjoint. 
Hence we have the following {\bf invertible frame rule}: 
\begin{center}
$\Pi \land \sigma*\Sigma \vdash \{\exists\Vec{y_i}(\Pi_i \land \sigma*\Sigma_i)\}_{i \in I}$
\ 
$\Longleftrightarrow$
\ 
$\Pi \land \Disj(\sigma,\Sigma) \land \Sigma \vdash \{\exists\Vec{y_i}(\Pi_i \land \Sigma_i)\}_{i \in I}$
\end{center}
We consider this invertible frame rule as a rewriting rule from the left-hand side to the right-hand side. 
By applying this rewriting rule as much as possible, 
the given entailment can be rewritten to entailments 
with a smaller number of $*$. 

This procedure has a great effect on efficiency improvement 
of our decision procedure, 
since reducing the number of $*$ highly contributes to reduce 
the sizes and the number of sorted entailments that are 
generated by the decomposition. 

\section{Implementation and Experiments}

This section explains our tool $\Checker$, 
which is the implementation of our decision procedure. 

\subsection{Entailment Checker $\Checker$}

The behavior of $\Checker$ is based on the decision procedure discussed in the previous sections.
It consists of the following three parts: 

(1) The optimizing part,
which reduces the size of a given entailment by the invertible frame rule ($\F$).
It also checks satisfiability of the antecedent of the entailment.
$\Checker$ immediately answers ``valid'' if the antecedent is unsatisfiable. 

(2) The decomposing part, 
which decomposes the reduced entailment 
into several sorted entailments, that is, 
the given unsorted entailment 
$\varphi \vdash \{\phi_i\}_{i \in I}$ is decomposed into sorted entailments 
$\Tilde{\varphi'} \vdash \{\Tilde{\phi'}\ |\ i \in I, \phi'\in Perm(\phi_i) \}$, 
where $\varphi' \in Perm(\varphi)$. 
The correctness of this part is guaranteed by Lemma~\ref{lemma:split_sorted}. 
After decomposition, redundant parts are reduced by the unsatisfiability checking ($\U$). 

(3) The translating part, 
which translates sorted entailments into Presburger formulas
according to the translation $P$
given in the section~\ref{sec:decidability}. 
The theoretical correctness of this part is guaranteed by 
Theorem~\ref{thm:correct}. 

(4) The checking part, in which the SMT-solver $\ZZZ$~\cite{Z3} is invoked to check validity of the generated Presburger formula. 

The current version of $\Checker$ is written in about 3900 lines of OCaml codes
(360 lines for the decomposing part and some optimization, 
1900 lines for the translating part, 
and 1200 lines for the checking part).

The improvements $\U$ and $\F$ mentioned above are optional, that is,
$\Checker$ has options which changes its behavior with (or without) them. 

\subsection{Experiments and Evaluation}

This subsection reports on the performance of $\Checker$. 
As far as we know, there is no suitable benchmark of entailments with arrays. 
So we automatically generated 120 entailments by using our analyzer of the C language. 
Each entailment has a single conclusion and is of small size 
(there are 1 or 2 separating conjunctions on each side).
We call this set of entailments $\Base$. 
40 out of 120 are entailments with only the points-to predicate (called the group Pto). 
Another 40 are entailments with only arrays (called the group Array). 
The rest 40 entailments contain both the points-to and array predicates (called the group Mix).
In each case, half of the entailments are valid. 

Then we automatically produced the following sets of entailments from $\Base$. 
All experimental files can be found in the web~\footnote{\texttt{https://github.com/DaisukeKimura/slar}}. 

- $\SingleFrame{n}$\ ($n = 2,3$) : 
A set of single-conclusion entailments produced by putting a frame of size $n$ to the entailments of $\Base$. 
The frames are chosen to keeping the grouping, that is,
frames of the points-to predicate are used for Pto,
frames of arrays are used for Array, and 
random frames are used for Mix; 

- $\SingleNFrame{n}$\ ($n = 2,3$) : 
A set of single-conclusion entailments produced by putting different spatial predicates of length $n$ to each side of the entailments of $\Base$. 
The spatial predicates are chosen to keeping the grouping; 

- $\Multi$ : 
A set of multi-conclusion entailments with at most 3 disjuncts. 
They are produced by adding extra conclusions to the entailments of $\Base$. 

In order to evaluate the effect of our improvement discussed in the previous subsection,
we provided the options that change whether $\Checker$ uses the unsatisfiability checking ($\U$) and the invertible frame rule ($\F$). 

For each categories $\Base$, $\SingleFrame{n}$, $\SingleNFrame{n}$ and $\Multi$, 
we executed our tool with (or without) the options of $\U$ and $\F$, 
and recorded its execution time (with timeout $300$ sec). 
Our PC environment is MacOS X machine with Intel Core i5 3.1GHz processor and 8GB memory. 

The results summarized in the tables of Fig.~\ref{fig:evaluation}, where 
$\U+\F$ means that the both options of the invertible frame rule and of the unsatisfiability checking are turned on.
$\U$ and $\F$ means that only corresponding option is used. 
None means that none of them are used. 
Each table shows the number of entailments (out of $120$) whose solved time satisfy the time condition displayed on the top of the table. 
For example, the table of $\Base$ shows that $111$ entailments are solved in less than $0.1$ sec by using both options ($\U+\F$). 

\begin{figure}[t]
\begin{center}
\scalebox{0.7}{
\begin{tabular}{|ccccc|c|}
\hline
\multicolumn{6}{|c|}{\Base}
\\
\hline
&&&&&\ time out
\\
&\ $<0.1$s&\ $<1$s&\ $<10$s&\ $<300$s&
($300$s)
\\
\hline
$\U+\F$&$111$&$120$&$120$&$120$&$0$
\\
$\U$&$103$&$120$&$120$&$120$&$0$
\\
$\F$&$106$&$120$&$120$&$120$&$0$
\\
None&$82$&$116$&$120$&$120$&$0$
\\
\hline
\end{tabular}
}
\qquad
\scalebox{0.7}{
\begin{tabular}{|ccccc|c|}
\hline
\multicolumn{6}{|c|}{\Multi}
\\
\hline
&&&&&\ time out
\\
&\ $<0.1$s&\ $<1$s&\ $<10$s&\ $<300$s&
($300$s)
\\
\hline
$\U+\F$&$57$&$108$&$120$&$120$&$0$
\\
$\U$&$45$&$106$&$120$&$120$&$0$
\\
$\F$&$24$&$101$&$115$&$120$&$0$
\\
None&$23$&$101$&$111$&$118$&$2$
\\
\hline
\end{tabular}
}
\end{center}
\begin{center}
\scalebox{0.7}{
\begin{tabular}{|ccccc|c|}
\hline
\multicolumn{6}{|c|}{\SingleFrame{2}}
\\
\hline
&&&&&\ time out
\\
&\ $<0.1$s&\ $<1$s&\ $<10$s&\ $<300$s&
($300$s)
\\
\hline
$\U+\F$&$101$&$118$&$120$&$120$&$0$
\\
$\U$&$78$&$119$&$120$&$120$&$0$
\\
$\F$&$97$&$116$&$120$&$120$&$0$
\\
None&$31$&$68$&$88$&$120$&$0$
\\
\hline
\end{tabular}
}
\qquad
\scalebox{0.7}{
\begin{tabular}{|ccccc|c|}
\hline
\multicolumn{6}{|c|}{\SingleFrame{3}}
\\
\hline
&&&&&\ time out
\\
&\ $<0.1$s&\ $<1$s&\ $<10$s&\ $<300$s&
($300$s)
\\
\hline
$\U+\F$&$81$&$107$&$118$&$120$&$0$
\\
$\U$&$37$&$95$&$120$&$120$&$0$
\\
$\F$&$80$&$105$&$116$&$119$&$1$
\\
None&$14$&$66$&$84$&$106$&$14$
\\
\hline
\end{tabular}
}
\end{center}
\begin{center}
\scalebox{0.7}{
\begin{tabular}{|ccccc|c|}
\hline
\multicolumn{6}{|c|}{\SingleNFrame{2}}
\\
\hline
&&&&&\ time out
\\
&\ $<0.1$s&\ $<1$s&\ $<10$s&\ $<300$s&
($300$s)
\\
\hline
$\U+\F$&$79$&$119$&$120$&$120$&$0$
\\
$\U$&$71$&$119$&$120$&$120$&$0$
\\
$\F$&$81$&$117$&$120$&$120$&$0$
\\
None&$31$&$80$&$111$&$119$&$1$
\\
\hline
\end{tabular}
}
\qquad
\scalebox{0.7}{
\begin{tabular}{|ccccc|c|}
\hline
\multicolumn{6}{|c|}{\SingleNFrame{3}}
\\
\hline
&&&&&\ time out
\\
&\ $<0.1$s&\ $<1$s&\ $<10$s&\ $<300$s&
($300$s)
\\
\hline
$\U+\F$&$34$&$102$&$119$&$120$&$0$
\\
$\U$&$33$&$87$&$118$&$120$&$0$
\\
$\F$&$39$&$97$&$117$&$120$&$0$
\\
None&$14$&$64$&$99$&$116$&$4$
\\
\hline
\end{tabular}
}
\end{center}
\caption{Experimental Results}
\label{fig:evaluation}
\end{figure}

\begin{figure}[h]
  \begin{center}
    \includegraphics[width=7.5cm]{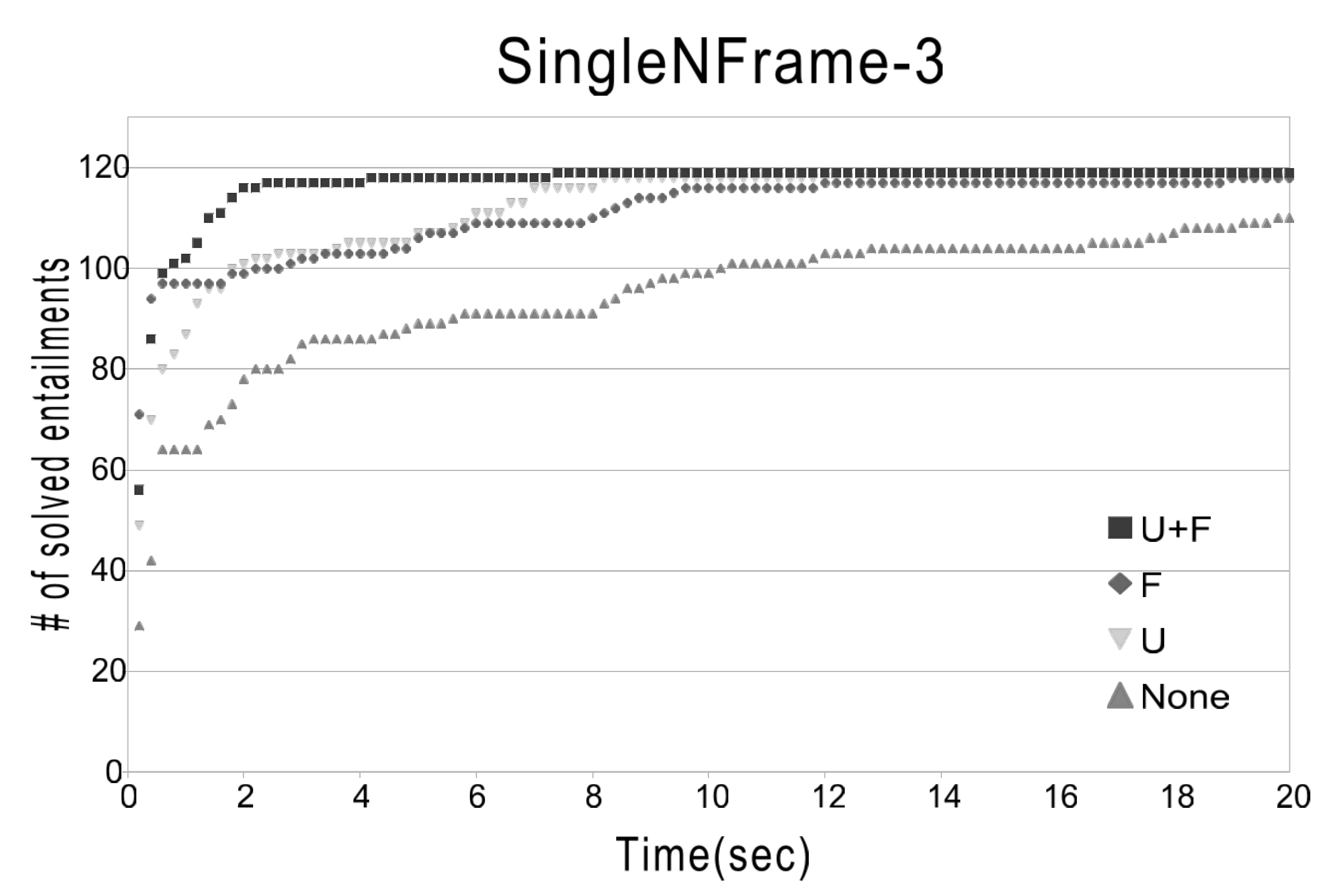}
  \end{center}
  \caption{Experimental Result of SingleNFrame-3}
  \label{fig:plot}
\end{figure}

For $\Base$, almost the entailments (111 out of 120) are answered within $0.1$ sec if both options are used. 
The rest 9 entailments are solved within $0.2$ sec. 
All are answered within $2.1$ sec without the acceleration options. 
Comparing $\U+\F$ and None, our tool becomes up to $15$ times, average $4$ times faster. 

The categories $\SingleFrame{n}$ ($n = 2,3$) show the effect of using the invertible frame rule, because entailments of these categories are in the form for which the invertible frame rule can be applied.
In $\SingleFrame{3}$, the cases of $\U+\F$ are about 10 times faster on average than that of $\U$ (without the invertible frame rule). Sometimes the tool becomes more than $200$ times faster! 

The categories $\SingleNFrame{n}$ ($n = 2,3$) are intended to limit use of the invertible frame rule. In the case of $\U+\F$ of $\SingleNFrame{3}$, almost entailments (102 out of 120) are solved within $1$ sec.

Checking the category $\Multi$ is mainly accelerated by the unsatisfiability checking. This is because unsatisfiable disjuncts of the succedent are eliminated in early stage of the decision procedure. The cases of $\U+\F$ are about $8$ times faster on average than that of $\F$ (without the unsatisfiability checking). 

In the case of $\U+\F$ of each category, the average time of the group Pto is less than $0.3$ sec, the average time of Array is less than $0.6$ sec, the average time of Mix is less than $0.7$ sec. 
The average time about invalid entailments is faster than that of valid entailments. It is because that our procedure immediately answers ``invalid'' when it finds an invalid decomposed entailment. 

From the results of $\Base$, 
our tool works very quickly for small entailments (the number of separating conjunctions on each side is less than or equal to 2).
However, as we expected, it becomes slower for entailments with greater sizes. 
Hence it is quite important to reduce the sizes of entailments. 
The results of $\SingleFrame{3}$ and $\SingleNFrame{3}$ shows that 
the invertible frame rule greatly contributes for reducing sizes and improving efficiency. 
The effect of our improvement remarkably appears when sizes of
entailments are large. In this experiment the effect worked well
in the case of $\SingleNFrame{3}$ (Fig.~\ref{fig:evaluation}). 

\section{Conclusion and Future Work}

In this paper we investigated the separation logic with arrays, and 
showed decidability of its entailment problem under the size condition. 
We also implemented our decision procedure that checks validity of entailments. 
The performance of our algorithm deeply depends on the number of separating conjunctions. 
So it is quite important to reduce their number. From the results of the experiments, 
we are convinced that the invertible frame rule and the undecidability checking work well 
and contribute for improving the entailment checking tool. 

Currently we have put a condition for proving correctness of the decision procedure. 
However our tool also seems to work well for entailments that are out of the condition. 
So we conjecture that correctness of the decision procedure also can be shown without the condition. 

Our algorithm is still inefficient since the performance of our tool slows sharply as the number of separating conjunction increases. 
One possible way is to split the given entailment into some smaller entailments, 
that is, split $\Pi_1\land \Sigma_1*\Sigma'_1 \vdash \Pi_2 \land \Sigma_2 * \Sigma'_2$ 
into $\Pi_1\land \Sigma_1 \vdash \Pi_2 \land \Sigma_2$ and $\Pi_1\land \Sigma'_1 \vdash \Pi_2 \land \Sigma'_2$. 
Of course the difficult point of this approach is to find the correct splitting point. 
However the memory model of the C language can be considered as the set of pairs of the form $(p,n)$, 
where $p$ is an identifier that indicates a domain and $n$ is an offset integer~\cite{Sekiguchi04}.  
If we carefully translate C codes into formulas of separation logic without losing the information of domains, 
it would be reasonable to choose a boundary of two different domains as a splitting point.

\newpage
\section*{Appendix}

\subsection*{Proof of Lemma~\ref{lemma:correct}}

\setcounter{lemma}{1}

\begin{lemma}
{\rm (1)}
Assume $s,h \models \hat \Pi \land \hat \Sigma$.
Suppose $P'(\Pi,\Sigma,S)$ appears in the unfolding of $P'(\hat\Pi, \hat\Sigma, \hat S)$.
Then

$
s,h|_{\Dom(s,\Sigma)} \models P'(\Pi,\Sigma,S)
$
iff
$
s,h|_{\Dom(s,\Sigma)} \models \Pi \land \Sorted(\Sigma) \imp \Lor\Tilde S.
$

{\rm (2)}
$
\forall sh(s,h \models \Pi\land\Tilde\Sigma \imp s,h \models \exists \Vec yP'(\Pi,\Sigma,S))
$
iff
$
\Pi\land\Tilde\Sigma \models \exists \Vec y\Lor\Tilde S.
$

{\rm (3)} $\models \neg(\Pi \land \Sorted(\Sigma)) \imp P(\Pi,\Sigma,S)$.

{\rm (4)}
$
\forall sh(s,h \models \Pi \land \Tilde\Sigma \imp s,h \models 
\forall \Vec z\exists \Vec yP(\Pi,\Sigma,S))
$
iff
$
\models \forall \Vec z\exists \Vec yP(\Pi,\Sigma,S).
$

\end{lemma}
\noindent\textbf{Proof of Lemma~\ref{lemma:correct}~(1)}. \quad
This is shown by induction on the steps $\eqDef$.
Consider cases according to the definition of $P'$.

\textit{Case 1} ($\mapsto\mapsto$-case):
\begin{align*}
P'&(\Pi,t \mapsto u * \Sigma,\{( \Pi_i, t_i \mapsto u_i * \Sigma_i) \}_{i \in I }) 
\\
&\eqDef P'(\Pi \land t < \Sigma,\Sigma, \{( \Pi_i \land t=t_i \land u=u_i \land t_i < \Sigma_i,\Sigma_i) \}_{i \in I}).
\end{align*}

Let $h_1=h|_{\Dom(s,t \mapsto u * \Sigma)}$,
$h_2=h|_{\Dom(s,\Sigma)}$. 
Then $h_1=\{(s(t),h(s(t)))\}+h_2$. 
It is enough to show
\begin{equation}\label{eq_1}
s,h_1 \models 
\Pi \land \Sorted(t \mapsto u * \Sigma) \imp \Lor_{i\in I} \Pi_i\land (t_i \mapsto u_i * \Sigma_i)^\sim
\end{equation}
iff
\begin{equation}\label{eq_2}
s,h_2 \models 
\Pi \land t < \Sigma \land \Sorted(\Sigma) \imp \Lor_{i\in I} \Pi_i\land t=t_i \land u=u_i \land t_i < \Sigma_i \land \Sigma_i^\sim.
\end{equation}

The only-if part.
Assume (\ref{eq_1}) and 
the antecedent of (\ref{eq_2}). 
Then the antecedent of (\ref{eq_1}) holds, since they are equivalent.
Then the succedent of (\ref{eq_1}) is true for $s,h_1$.
Hence the succedent of (\ref{eq_2}) is true for $s,h_2$.

The if part.
Assume (\ref{eq_2}) and 
the antecedent of (\ref{eq_1}).
Then the antecedent of (\ref{eq_2}) holds, since they are equivalent.
Then the succedent of (\ref{eq_2}) is true for $s,h_2$.
Hence the succedent of (\ref{eq_1}) is true for $s,h_1$.

\textit{Case 2} ({\bf Arr}$\mapsto$-case):
\begin{align*}
P'(\Pi,\Array(t,t') * \Sigma,S) 
\eqDef&
P'(\Pi \land t'=t \land t \le \Sigma,\Sigma,S)
\\
&\land
P'(\Pi \land t'>t,t \mapsto [t] * \Array(t+1,t') * \Sigma,S) 
\end{align*}

Let $h_3 = h|_{\Dom(s,\Array(t,t') * \Sigma)}$,
$h_4=h|_{\Dom(s,t\mapsto [t] * \Sigma)}$, and 
$h_5=h|_{\Dom(s,t \mapsto [t] * \Array(t+1,t') * \Sigma)}$. 
It is enough to show 
\begin{equation}\label{eq_3}
s,h_3 \models 
\Pi \land \Sorted(\Array(t,t') * \Sigma) \imp \Lor \Tilde S
\end{equation}
is equivalent to the conjunction of the following two clauses: 
\begin{eqnarray}
  \lefteqn{
    s,h_4 \models
    \Pi \land t'=t \land t < \Sigma \land \Sorted(\Sigma) \imp \Lor \Tilde S
    \quad\hbox{and}
  }
  \label{eq_4}
  \\
  \lefteqn{
    s,h_5 \models
    \Pi \land t'>t \land \Sorted(t \mapsto [t] * \Array(t+1,t') * \Sigma)
    \imp \Lor \Tilde S.
    }
  \label{eq_5}
\end{eqnarray}

\textit{Case 2.1}: the case of $s(t)=s(t')$.

We note that $h_3 = h_4$. 
The antecedent of (\ref{eq_4}) is equivalent to
the antecedent of (\ref{eq_3}). 
(\ref{eq_5}) is true since $s(t)=s(t')$.
Hence (\ref{eq_3}) and $(\ref{eq_4})\land(\ref{eq_5})$ are equivalent.

\textit{Case 2.2}: the case of $s(t')>s(t)$. 

We note that $h_3=h_5$.
The antecedent of (\ref{eq_5}) is equivalent to
the antecedent of (\ref{eq_3}).
(\ref{eq_4}) is true since $s(t')>s(t)$. 
Hence (\ref{eq_3}) and $(\ref{eq_4})\land(\ref{eq_5})$ are equivalent.

\textit{Case 3} ($\mapsto${\bf Arr}-case):
\begin{align*}
P'(\Pi,&t \mapsto u * \Sigma,\{(\Pi_i,\Array(t_i,t_i') * \Sigma_i)\} \cup S)
\\
\eqDef&
P'(\Pi \land t_i'=t_i,t \mapsto u * \Sigma,\{(\Pi_i,t_i \mapsto u * \Sigma_i)\} \cup S)
\\ 
&\land
P'(\Pi \land t_i'>t_i,t \mapsto u * \Sigma,\{(\Pi_i,t_i \mapsto u * \Array(t_i+1,t_i') * \Sigma_i)\} \cup S)
\\
&\land
P'(\Pi \land t_i'<t_i,t \mapsto u * \Sigma,S)
\end{align*}
This case is proved by showing the following claim, 
which is shown similarly to the claim of Case 2.
Let $h' = h|_{\Dom(s,t \mapsto u * \Sigma)}$. Then 
\[
s,h'\models 
\Pi \land \Sorted_L \imp 
\Pi_i \land (\Array(t_i,t_i') * \Sigma_i)^\sim \lor \Lor \Tilde S
\]
is equivalent to the conjunction of the following three clauses: 
\\
\begin{tabular}{ll}
$s,h' \models$&
$\Pi \land t_i'=t_i\land \Sorted_L$
$\imp 
\Pi_i \land (t_i \mapsto u * \Sigma_i)^\sim \lor \Lor \Tilde S$
\\
$s,h' \models$&
$\Pi \land t_i'>t_i\land \Sorted_L$
$\imp \Pi_i \land (t_i \mapsto u * \Array(t_i+1,t_i') * \Sigma_i)^\sim \lor \Lor \Tilde S$, 
\\
$s,h' \models$&
$\Pi \land t_i'<t_i\land \Sorted_L \imp \Lor \Tilde S$, 
\end{tabular}
\\
where $\Sorted_L$ is an abbreviation of $\Sorted(t \mapsto u * \Sigma)$. 

\textit{Case 4} ({\bf (ArrArr)}-case): 
Consider that
$P(\Pi,\Array(t,t') * \Sigma,\{( \Pi_i,\Array(t_i,t_i') * \Sigma_i)\}_{i \in I})$
is defined by the conjunction of 
\[
  P\left(
  \begin{array}{l}
    \Pi \land
    m=m_{I'}\land m<m_{I\setminus I'}
    \land t \le t' \land t' < \Sigma, \Sigma,
    \\
    \{( \Pi_i \land t_i+m < \Sigma_i, \Sigma_i)\}_{i \in I'}
    \cup
    \{( \Pi_i,\Array(t_i+m+1,t_i') * \Sigma_i)\}_{i \in I\setminus I'}
  \end{array}
  \right)
\]
for all $I'\subseteq I$ and 
\[
  P\left(
  \begin{array}{l}
    \Pi \land m'<m \land m'=m_{I'} \land m'<m_{I\setminus I'},
    \Array(t+m'+1,t') * \Sigma,
    \\
    \{( \Pi_i \land t_i+m' < \Sigma_i, \Sigma_i )\}_{i \in I'}
    \cup
    \{( \Pi_i,\Array(t_i+m'+1,t_i') * \Sigma_i )\}_{i \in I\setminus I'}
  \end{array}
  \right)
  \]
for all $I' \subseteq I$ with $I' \neq \emptyset$, 
where $m$, $m_i$, and $m'$ are abbreviations of 
$t'-t$, $t_i'-t_i$, and $m_{\min I'}$, respectively.

Let 
$h_6 = h|_{\Dom(s,\Array(t,t') * \Sigma)}$, 
$h_7 = h|_{\Dom(s,\Sigma)}$, and 
$h_8 = h|_{\Dom(s,\Array(t+m'+1,t') * \Sigma)}$. 
It is enough to show 
\begin{equation}
\label{eq_6}
s,h_6 \models 
\Pi \land \Sorted(\Array(t,t') * \Sigma) \imp \Lor_{i\in I} \Pi_i \land (\Array(t_i,t_i') * \Sigma_i))^\sim
\end{equation}
is equivalent to the conjunction of the following
\begin{eqnarray}
  \lefteqn{
    s,h_7 \models
    \Pi\land m=m_{I'} \land m<m_{I\setminus I'}
    \land t \le t' \land t' < \Sigma \land \Sorted(\Sigma)
  }
  \nonumber
  \\
  \lefteqn{
    \imp\Lor_{i\in I'}
    \Pi_i \land t_i+m' < \Sigma_i\land\Sigma_i^\sim
    \ \vee\
    \Lor_{i\in I\setminus I'}\Pi_i\land (\Array(t_i+m'+1,t_i') * \Sigma_i)^\sim
  }  
  \label{eq_7}
\end{eqnarray}
for any $I' \subseteq I$, 
and
\begin{eqnarray}
  \lefteqn{
    s, h_8 \models
    \Pi\land m'<m \land m'=m_{I'}
    \land m'<m_{I\setminus I'}
    \land \Sorted(\Array(t+m'+1,t') * \Sigma)
  }  
  \nonumber  
  \\
  \lefteqn{
    \imp \Lor_{i \in I'}
    \Pi_i \land t_i+m' < \Sigma_i\land\Sigma_i^\sim
    \ \vee\
    \Lor_{i \in I-I'}\Pi_i\land(\Array(t_i+m'+1,t_i') * \Sigma_i)^\sim
  }
  \label{eq_8}
\end{eqnarray}
for any $I' \subseteq I$ with $I' \neq \emptyset$.

\textit{Case 4.1}: the case of $s \models m=m_{I'} \land m<m_{I\setminus I'}$ for some $I' \subseteq I$.

The antecedent of the conjunct of (\ref{eq_7}) with respect to $I'$ has the case condition.
All conjuncts of (\ref{eq_7}) other than this conjunct and 
conjuncts of (\ref{eq_8}) are true,
because their antecedents are false by the case condition.

Now we show the only-if part and the if part 
by using $h_6= h|_{\{s(t),s(t+1),\ldots,s(t')\}} + h_7$.

The only-if part: 
Assume (\ref{eq_6}) and the antecedent of the conjunct.
Then the antecedent of (\ref{eq_6}) holds, since they are equivalent.
Then the succedent of (\ref{eq_6}) is true for $s,h_6$.
Hence the succedent of the conjunct is true for $s,h_7$.

The if part:
Assume (\ref{eq_7}) and the antecedent of (\ref{eq_6}).
Then the antecedent of the conjunct holds, since they are equivalent.
Then the succedent of the conjunct is true for $s,h_7$.
Hence the succedent of (\ref{eq_6}) is true for $s,h_6$.

\textit{Case 4.2}: the case of  
$s \models m'<m \land m'=m_{I'} \land m'<m_{I\setminus I'}$ for some $I' \subseteq I$.
It is similar to Case 4.1 by using
$h_6 = h|_{\{s(t), \ldots, s(t+m') \}} + h_8$. 

\noindent\textbf{Proof of Lemma~\ref{lemma:correct}~(2)}.\quad
We first show the only-if part. 
Assume the left-hand side of the claim and 
$s,h \models \Pi \land \Tilde\Sigma$.
By the left-hand side, we obtain 
$s,h \models \exists \Vec yP'(\Pi,\Sigma,S)$.
Hence we have $s'$ such that $s',h \models P'(\Pi,\Sigma,S)$.
By (1),
$s',h \models \Pi \land \Sorted(\Sigma) \imp \Lor S$.
Thus $s',h \models \Lor S$. 
Finally we have $s,h \models \exists \Vec y\Lor S$.

Next we show the if part. 
Fix $s,h$.
Assume the right-hand side of the claim and 
$s,h \models \Pi \land \Tilde\Sigma$.
By the right-hand side, $s,h \models \exists \Vec y\Lor \Tilde S$ holds.
Hence we have $s'$ such that $s',h \models \Lor S$.
Then we obtain $s',h \models \Pi \land \Sorted(\Sigma) \imp \Lor \Tilde S$.
By (1), $s',h \models P'(\Pi,\Sigma,S)$ holds.
Finally we have $s,h \models \exists \Vec yP'(\Pi,\Sigma,S)$.

\noindent\textbf{Proof of Lemma~\ref{lemma:correct}~(3)}.\quad
It is shown by induction on the steps of $\eqDef$.

\noindent\textbf{Proof of Lemma~\ref{lemma:correct}~(4)}.\quad
We note that 
$\forall sh(s,h \models \Pi \land \Tilde\Sigma \imp s \models \forall \Vec z\exists \Vec yP(\Pi,\Sigma,S))$
is equivalent to
$\forall s(\exists h(s,h \models \Pi \land \Tilde\Sigma) \imp s \models \forall \Vec z\exists \Vec yP(\Pi,\Sigma,S))$.
Moreover it is equivalent to
$\forall s(s \models \Pi \land \Sorted(\Sigma) \imp s \models \forall \Vec z\exists \Vec yP(\Pi,\Sigma,S))$.
By using (3), it is equivalent to
$\forall s(s \models \forall \Vec z\exists \Vec yP(\Pi,\Sigma,S))$,
namely, $\models \forall \Vec z\exists \Vec yP(\Pi,\Sigma,S)$.
$\Box$

\end{document}